# Integration of Genetic Algorithms and Deep Learning for the Generation and Bioactivity Prediction of Novel Tyrosine Kinase Inhibitors

Ricardo Romero*[a]

**Abstract:** The intersection of artificial intelligence and bioinformatics has enabled significant advancements in drug discovery, particularly through the application of machine learning models. In this study, we present a combined approach using genetic algorithms and deep learning models to address two critical aspects of drug discovery: the generation of novel tyrosine kinase inhibitors and the prediction of their bioactivity. The generative model leverages genetic algorithms to create new small molecules with optimized ADMET (absorption, distribution, metabolism, excretion, and toxicity) and drug-likeness properties. Concurrently, a deep learning model is employed to predict the bioactivity of these generated molecules against tyrosine kinases, a key enzyme family involved in various cellular processes and cancer progression. By integrating these advanced computational methods, we demonstrate a powerful framework for accelerating the generation and identification of potential tyrosine kinase inhibitors, contributing to more efficient and effective early-stage drug discovery processes.

## 1 Introduction

The intersection of artificial intelligence and bioinformatics has heralded a new era in drug discovery and cheminformatics, where traditional methods are complemented and often surpassed by machine learning techniques. Among these, neural networks and genetic algorithms have emerged as powerful tools due to their unique characteristics and capabilities.

Neural networks, inspired by biological neural systems, excel at pattern recognition and complex data processing, making them invaluable for predicting molecular properties and activities. Genetic algorithms, which mimic natural selection processes, are adept at optimizing solutions in vast search spaces, such as those encountered in molecular design.

Generative models, particularly those based on deep learning architectures, have shown remarkable success in creating novel molecular structures. Variational autoencoders (VAEs) and generative adversarial networks (GANs) have been widely adopted for this purpose. For instance, Gómez-Bombarelli et al. demonstrated the use of VAEs to generate molecules with desired properties in a continuous latent space.[1]

Reinforcement learning (RL) has emerged as a powerful tool for optimizing molecular structures towards specific objectives. Jin et al. developed a graph convolutional policy network that learns to generate molecules with desired properties through RL, showcasing its effectiveness in generating molecules with high QED (quantitative estimate of drug-likeness) scores.[2]

The challenge of ensuring synthetic accessibility of generated molecules has been addressed by incorporating reaction-based approaches. Bradshaw et al. introduced a model that generates molecules using a set of known chemical reactions, ensuring that the proposed structures are synthetically viable.[3]

Recent work has also focused on multi-objective optimization in molecular generation. Blaschke et al. developed a method that combines genetic algorithms with deep learning to generate molecules optimized for multiple properties simultaneously, a crucial aspect in practical drug discovery scenarios.[4]

Despite their strengths, both approaches have limitations. Neural networks often require large datasets for training and can be prone to overfitting, while genetic algorithms may converge on local optima, rather than the global, and struggle with highly constrained problems. Nevertheless, their application in drug discovery has yielded promising results, particularly in areas such as virtual screening, quantitative structure-activity relationship (QSAR) modelling, and de novo drug design.[5]

In the realm of cheminformatics, these techniques have been instrumental in developing predictive models for various molecular properties, including bioactivity, toxicity, and pharmacokinetics.[6] The integration of neural networks and genetic algorithms with traditional cheminformatics approaches has led to more efficient and accurate computational methods for drug discovery.

[a] *Departamento de Ciencias Naturales, UAM Unidad Cuajimalpa.*
*Avenida Vasco de Quiroga 4871, Col. Santa Fe Cuajimalpa. Alcaldía Cuajimalpa de Morelos, C.P. 05348, Ciudad de México.*
*\*e-mail: rromero@cua.uam.mx*

*Supporting Information for this article is available at https://github.com/ricardo-romero-ochoa/bio-deep-learning*

This paper aims to address these challenges by presenting two models, one for generation and one for bioactivity classification of tyrosine kinase inhibitors, trained on a carefully curated dataset from the ChEMBL database.[7] Tyrosine kinases are enzymes that play a critical role in many cellular processes, and their inhibitors are vital in the treatment of various cancers. Our model leverages neural networks to predict the bioactivity of these inhibitors, providing a robust and accurate tool for drug discovery. Additionally, we introduce a generative model that uses genetic algorithms to produce new small molecules with optimal ADMET (absorption, distribution, metabolism, excretion, and toxicity) as well as druglikeness properties, thus addressing the need for efficient and effective drug design.

Our work aims to demonstrate the synergistic potential of neural networks and genetic algorithms in addressing two critical aspects of drug discovery: the identification of potent inhibitors for a specific target class and the generation of drug-like molecules with desirable pharmacokinetic profiles. By combining these approaches, we seek to contribute to the ongoing efforts to streamline and accelerate the drug discovery process.

## 2 Computational methods

### 2.1 Data Collection and Preprocessing

The dataset [8] for this study was sourced from the ChEMBL database, focusing on human tyrosine kinases as targets. Eleven specific tyrosine kinases were selected: ABL (Abelson tyrosine-protein kinase 1), EGFR (Epidermal growth factor receptor), PDGFR (Platelet-derived growth factor receptor), FGFR (Fibroblast growth factor receptor), MET (Hepatocyte growth factor receptor), VEGFR (Vascular endothelial growth factor receptor), KIT (Stem cell factor receptor), RET (Rearranged during transfection), JAK (Janus kinase), ALK (Anaplastic lymphoma kinase), and SRC (Proto-oncogene tyrosine-protein kinase).

These tyrosine kinases are implicated in various diseases, including cancer, and targeting them with specific inhibitors has shown therapeutic benefits in clinical settings. IC50 values, in nMol, for these targets were extracted from ChEMBL. The dataset underwent rigorous preprocessing to ensure data quality. This included removing entries with missing or zero IC50 values and those lacking SMILES representations or containing duplicate SMILES.

### 2.2 Bioactivity Classification

Compounds were classified based on their IC50 values using the following criteria:
- Inactive: IC50 ≥ 10,000 nM
- Active: IC50 ≤ 1,000 nM
- Intermediate: 1,000 nM < IC50 < 10,000 nM

To standardize the data, a maximum threshold of 10^9 nM was applied to the IC50 values, with all values equal to or exceeding this threshold set to 1. Subsequently, all IC50 values were converted to pIC50 (-log10(IC50)) for further analysis.

### 2.3 Molecular Descriptors, Toxicity and Drug-like Properties

Lipinski molecular descriptors were computed using the RDKit module in Python[9] and the DataWarrior software was employed to evaluate the toxicity profiles and druglikeness of the compounds in the dataset.

This methodology ensures a comprehensive and well-curated dataset of tyrosine kinase inhibitors, incorporating both bioactivity data and key molecular properties relevant to drug discovery and development.

### 2.4 Bioactivity classification deep learning model

A convolutional neural network (CNN) model was developed to classify the bioactivity of tyrosine kinase inhibitors. The model was implemented using Python, with key libraries including TensorFlow/Keras for deep learning,[10] RDKit for cheminformatics, and scikit-learn for data preprocessing and evaluation metrics.[11]

Data Preparation:

1. The tyrosine kinase dataset, containing SMILES representations of molecules and their pIC50 values, was loaded from a CSV file.
2. SMILES strings were converted to RDKit molecule objects, from which RDKit fingerprints were generated.
3. Bioactivity data was binarized: compounds with pIC50 ≥ 6 were labeled as active (1), and others as non-active (0).
4. Features consisted of four molecular descriptors and the RDKit fingerprints.
5. Data was standardized using scikit-learn's StandardScaler.
6. The dataset was split into training (80%) and test (20%) sets.

Model Architecture:

A 1D CNN was constructed with the following layers:

1. Two convolutional layers (32 and 16 filters, kernel size 3).
2. Max pooling layers after each convolution.
3. Dropout layers (30% dropout rate) for regularization.
4. A flatten layer.
5. A dense layer with 64 units and ReLU activation.
6. A final dense layer with sigmoid activation for binary classification.

The model was compiled using binary cross-entropy loss and the Adam optimizer, and then trained for up to 100 epochs with early stopping and learning rate reduction callbacks to prevent overfitting. A batch size of 32 was used. Model performance was assessed using several metrics:
1. Accuracy, precision, recall, and F1 score
2. Receiver Operating Characteristic (ROC) curve and Area Under the Curve (AUC)
3. Confusion matrix
4. 5-fold stratified cross-validation was performed to ensure robustness of the model's performance.
5. The trained model was saved in both Keras (.keras) and HDF5 (.h5) formats for future use.[12]

This approach allows for the classification of tyrosine kinase inhibitors based on their predicted bioactivity, potentially accelerating the drug discovery process by identifying promising compounds for further investigation.

### 2.5 Generative model from genetic algorithm

A genetic algorithm-based approach was implemented to generate novel tyrosine kinase inhibitors with optimized drug-like properties. The algorithm was developed in Python, utilizing the RDKit library for molecular manipulation and property calculation, the SELFIES (Self-Referencing Embedded Strings) library for molecular representation,[13] and pandas for data handling.[14]

The genetic algorithm process consisted of the following key steps:

1. Data Loading: A subset of the tyrosine kinase inhibitors, containing only active compounds with none of the toxicity features and druglikeness ≥ 1 was loaded, providing the initial pool of SMILES (Simplified Molecular Input Line Entry System) strings.
2. Molecular Representation: Molecules were represented using the SELFIES format, which helps in generating valid structures. Conversion functions were implemented to translate between SMILES and SELFIES representations.
3. Fitness Function: A custom fitness function was designed to evaluate the quality of generated molecules. This function incorporated:
   - Quantitative Estimate of Drug-likeness (QED) [15]
   - LogP (octanol-water partition coefficient)
   - Topological Polar Surface Area (TPSA)

The fitness function aimed to maximize QED while keeping LogP and TPSA close to target values (3.0 and 75.0, respectively). This multi-objective optimization approach balances druglikeness with specific physicochemical properties.

4. Genetic Operators:
   - Mutation: Random alterations to the SELFIES string, with a mutation rate of 0.01.
   - Crossover: Single-point crossover between two parent molecules, with a crossover rate of 0.7.
   - Selection: Tournament selection with a tournament size of 5.

5. Algorithm Execution: The genetic algorithm ran for 50 generations with a population size of 100. In each generation:
   - New offspring were created through crossover and mutation.
   - Fitness was evaluated for all new individuals.
   - The population was sorted based on fitness.

6. Molecule Generation: The algorithm was executed 100 times with seed number 42 to assure reproducibility, and to generate a diverse set of optimized molecules.

7. Output: The generated molecules were filtered for toxicity and druglikeness, the 2D structures validated and the 3D structures optimized. Stereochemistry and tautomers were also calculated.

This approach allows for the exploration of a vast chemical space while maintaining the core properties that make molecules suitable as tyrosine kinase inhibitors. By optimizing for druglikeness and key physicochemical properties simultaneously, the algorithm aims to generate molecules that are not only potentially active but also have favorable ADMET profiles.

## 3 Results

Out of the one hundred generated molecules, five passed all the toxicity, druglikeness and validation filters. They are given in Figures 1-5 and their properties in Tables I and II.

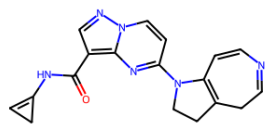

**Figure 1.** 2D structure of generated molecule 1, with SMILES C1C=NC=CC2=C1CCN2C=3C=CN4N=CC(C(=O)NC=5CC=5)=C4N=3

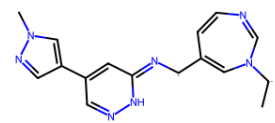

**Figure 2.** 2D structure of generated molecule 2, with SMILES C[C@@H1]N1C=NC=CC([C@@H1]N=C2C=C(C=3C=NN(C)C=3)C=NN2)=C1

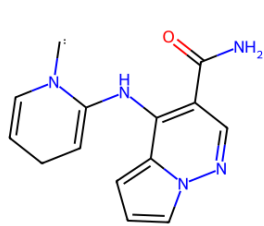

**Figure 3.** 2D structure of generated molecule 3, with SMILES [C@@H1]N1C=CCC=C1NC2=C(C(N)=O)C=NN3C=CC=C23

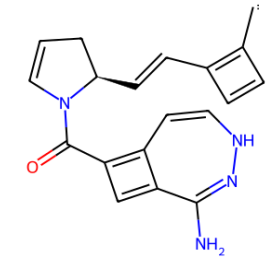

**Figure 4.** 2D structure of generated molecule 4, with SMILES NC1=N[NH1]C=CC2=C(C(=O)N3C=CC[C@H1]3C=CC4=C([C@H1])C=C4)C=C12

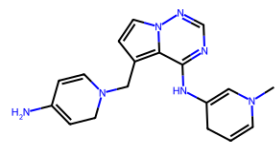

**Figure 5.** 2D structure of generated molecule 5, with SMILES CN1C=CCC(NC2=NC=NN3C=CC(CN4C=CC(N)=CC4)=C23)=C1

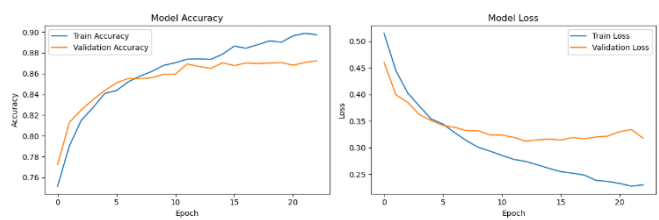

**Table I. Generated Molecules Properties.** The fitness score was provided by the genetic algorithm script, while solubility, druglikeness and drug-score were calculated with Osiris Property Explorer. Synthetic accessibility scores (SA) were obtained from the SwissADME server.[16] All the molecules were classified as active by the bioactivity deep learning model.

| Molecule | Fitness | Solubility | Drug likeness | Drug Score | SA |
|---|---|---|---|---|---|
| 1 | 0.93 | -4.53 | 7.68 | 0.76 | 3.78 |
| 2 | 0.92 | -2.13 | 6.97 | 0.93 | 3.97 |
| 3 | 0.86 | -3.56 | 5.69 | 0.87 | 3.06 |
| 4 | 0.91 | -2.87 | 5.02 | 0.89 | 4.04 |
| 5 | 0.88 | -3.2 | 4.93 | 0.87 | 3.81 |

**Figure 6. Training and Validation Accuracy and Loss Over Epochs.** The plot shows the performance of the bioactivity classification model during training, as measured by accuracy and loss across epochs. Overall, the plots suggest that the model is learning effectively, but there may be early signs of overfitting as the validation loss begins to rise slightly while accuracy levels off, hence the need for early stopping.

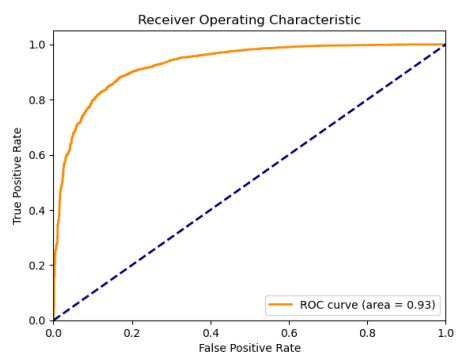

**Figure 7. Receiver Operating Characteristic (ROC) Curve.** The ROC curve demonstrates a strong performance, as it stays well above the diagonal line, with an AUC (Area Under the Curve) of 0.93. This high AUC value indicates that the model has excellent discriminatory ability, effectively distinguishing between the positive and negative classes.

**Table II. Generated Molecules Descriptors.** Lipinski[17] descriptors: molecular weight (MW) in g/mol, water octanol partition function (LogP), and number of H bonds acceptors/donors. Topological surface area (TPS) is given in Å². The compounds also pass the Ghose,[18] Veber,[19] Egan,[20] Muegge,[21] and lead likeness filters, and show neither Brenk[22] nor PAINS[23] (Pan Assays Interference Structures) structural alerts. All the values were obtained from the SwissADME server.

| Molecule | MW | LogP | NumH acceptors | NumH donors | TPSA |
|---|---|---|---|---|---|
| 1 | 332.37 | 2.20 | 4 | 1 | 74.89 |
| 2 | 307.36 | 1.80 | 4 | 1 | 74.46 |
| 3 | 267.29 | 1.57 | 2 | 2 | 75.66 |
| 4 | 316.36 | 2.96 | 2 | 2 | 75.01 |
| 5 | 335.42 | 2.00 | 2 | 2 | 74.72 |

For the bioactivity classification model, training and validation accuracy/loss curves, ROC curve, and confusion matrix are given in Figures 6-8 for visual interpretation of the model's performance. Model's metrics are provided in Table III.

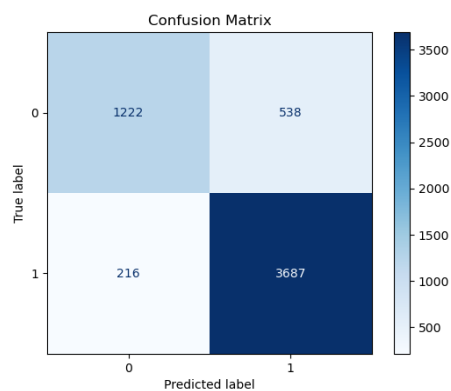

**Figure 8. Confusion Matrix of Model Predictions.** This confusion matrix visualizes the performance of the bioactivity classification model by comparing the predicted labels (x-axis) to the true labels (y-axis). The color intensity in the matrix indicates the number of instances, with darker shades representing higher counts. This matrix shows that the model performs well, with a higher number of true positives and true negatives compared to false positives and false negatives, suggesting strong overall classification accuracy.

# 4 Discussion and concluding remarks

The results of this study highlight the synergistic potential of combining genetic algorithms with deep learning models in drug discovery. The generative model effectively produced novel tyrosine kinase inhibitors with favourable pharmacokinetic profiles, addressing a key challenge in the early stages of drug design. Additionally, the bioactivity prediction model provided accurate assessments of the generated molecules, demonstrating its utility in virtual screening and candidate selection.

The integration of these approaches offers a comprehensive framework for the design and evaluation of new drug candidates, significantly reducing the time and resources typically required in drug discovery. While the study focused on tyrosine kinase inhibitors, the methodology is generalizable and can be applied to other drug targets, making it a versatile tool in cheminformatics.

However, it's important to acknowledge the limitations of our approach. The accuracy of the bioactivity prediction model is contingent on the quality and diversity of the training data and may not generalize well to novel chemical scaffolds. Similarly, while the genetic algorithm ensures better exploration of the chemical space, it may sometimes converge on local optima rather than global ones.

Future work should focus on validating these computational predictions through experimental testing, refining the models based on new data, and potentially expanding the approach to other target classes beyond tyrosine kinases. Additionally, incorporating more sophisticated measures of synthetic accessibility and integrating reaction-based approaches could further enhance the practical applicability of the generated molecules.

In conclusion, this study demonstrates the effectiveness of combining genetic algorithms and deep learning for both the generation and bioactivity prediction of novel drug candidates. This integrated approach has the potential to streamline the drug discovery process, providing a powerful tool for the identification and optimization of new therapeutic agents.